\documentclass{PoS}
\usepackage{amsmath,epsfig,cite,graphics}
\usepackage{macros_static}
\usepackage{wrapfig}

\title{Running of the SF-coupling with four massless flavours
}

\ShortTitle{Running of the SF-coupling with four massless flavours
}

\author{\LALPHA \hfill
        \onecol{4.0cm}{\vspace{-1.5cm}\it DESY 10-192 \\ SFB/CPP-10-102
          \\ HU-EP-10/72
          \vspace{-2.1cm}
        }}

\author{\speaker{Rainer Sommer}\\ %\thanks{A footnote may follow.}        
        NIC, DESY, Platanenalle 6, 15738 Zeuthen, Germany\\
        E-mail: \email{rainer.sommer@desy.de}\\
        }
\author{Fatih Tekin, Ulli Wolff\\
        Humboldt University, Newtonstr.~15, 12489 Berlin, Germany
       }

\abstract{We discuss the status of different determinations of $\alpha_\mrm{s}$,
motivating a precise and reliable computation from lattice
QCD. In order to suppress perturbative errors, the non-perturbative
computation has to reach high energy scales $\mu$. Such results already
exist in the SF-scheme for $\Nf=0,2$ \cite{mbar:pap1,alpha:nf2}
and  $\Nf=3$ \cite{alpha:nf3}. We recently added the running with
four massless flavours in a range of $\alpha$ from  about 0.07 to
0.3 . It is based on our recent determination of the Sheikholeslami
Wohlert coefficient in the four-flavour theory.}

\FullConference{The XXVIII International Symposium on Lattice Field Theory, 
Lattice2010\\
		June 14-19, 2010\\
		Villasimius, Italy}

\begin{document}

%%%%%%%%%%%%%%%%%%%%%%%%%%%%%%%%%%%%%%%%%%%%%%%%%%%%%%%%%%%%
\begin{wrapfigure}{r}{0.4\textwidth}
\vspace{-1.0cm}
\begin{center}
\includegraphics[width=0.4\textwidth]{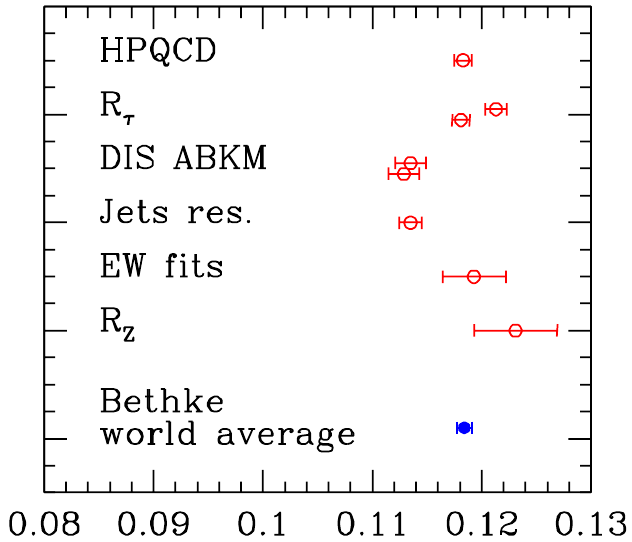}
\end{center}
\vspace{-0.5cm}
\caption[]{\label{f:summary} Some recent 
precise numbers for $\alpha_\mrm{s}(M_\mrm{Z})$ 
in the $\msbar$ scheme. From top
to bottom: lattice QCD by HPQCD \protect\cite{alpha:HPQCD},
hadronic $\tau$ decays ($R_\tau$) \protect\cite{alpha:tau:beneke},
deep inelastic scattering \protect\cite{alpha:ABKM}, 
thrust \protect\cite{alpha:AFHMS},
global electroweak fit \protect\cite{alpha:EW},
hadronic decays of the Z ($R_Z$) \protect\cite{rz:exp,rz:4-loop} and 
the world average of
\protect\cite{alpha:bethke09}.
}
\end{wrapfigure}
%%%%%%%%%%%%%%%%%%%%%%%%%%%%%%%%%%%%%%%%%%%%%%%%%%%%%%%%%%%%
\section{Introduction}
The strong coupling, $\alphas$, represents a fundamental parameter 
of the strong interactions. Its scale dependence in a suitable 
renormalization scheme teaches us about the connection between
the strongly non-perturbative and the dominantly perturbative 
regions of QCD. Its value at high energy  
is needed for phenomenology, for example for the prediction of
Higgs production cross sections for the LHC.
The uncertainty on $\alphas$ is {\em not} negligible in this 
context~\cite{alpha:ABKM}. 

For this conference we had compiled a plot of precise
determinations of $\alpha\equiv\alpha_\mrm{s}$ with the 
renormalization scale set to the mass of the Z-boson  and 
in the $\msbar$ scheme. It is shown in \fig{f:summary}. 
We do not claim completeness but rather want to illustrate 
that the many determinations do not agree well within the 
estimated uncertainties. 
%Note that the
%ALPHA collaboration has so far not given any value of 
%$\alpha$ in the physical theory with 
%all quarks. 

Indeed, the spread of results in \fig{f:summary} is not that 
surprising, since a precision determination of $\alpha_\mrm{s}(M_\mrm{Z})$
is difficult and so far compromises on various sources of errors 
had to be made despite an ever increasing sophistication
in the analysis. Sources of errors are:
\bi
   \vspace*{-1mm}
   \item Low energy: most determinations are not done
     from a process with an energy scale of order $M_\mrm{Z}$
     but at significantly lower energies and are then
     evolved perturbatively to $\mu=M_\mrm{Z}$. A 
     prominent example is the determination from $\tau$-decays,
     labeled $R_\tau$. 
   \vspace*{-1mm}\item Sophisticated analysis involving simultaneous fits to
     $\alpha_\mrm{s}(\mu)$ and non-perturbative parameters of
     QCD, such as structure functions~\cite{alpha:ABKM} and 
     parameters of SCET~\cite{alpha:AFHMS}.
   \vspace*{-1mm}\item Global fits to many processes (EW fits) \cite{alpha:EW},
     which of course means that the correctness and 
     theoretical mastering of the standard model enters in detail.
   \vspace*{-1mm}\item The use of bare, unrenormalized, perturbation theory 
     of Wilson loops at the cutoff scale in the lattice
     determination \cite{alpha:HPQCD} shown in the figure.
     We discuss this further below.
   \vspace*{-1mm}\item None of these applies to the extraction from the 
     hadronic cross section on the Z-peak, $R_\mrm{Z}$, but this is
     experimentally more difficult, resulting in a larger quoted error.
\ei
\vspace*{-1mm}The present world average by S. Bethke is dominated by the lattice
determination \cite{alpha:HPQCD}. We find this worrying due to
several reasons. First there is the 
use of bare perturbation theory. 
There the problem is that successive terms do not show
a convincing `convergence' pattern.
Either there are large coefficients of 
higher order terms (expansion in the original bare coupling) 
or the (modified) coupling itself ``lives'' at a quite low energy scale 
where it is large (``tadpole improved'' coupling, coupling in
``potential scheme''). In either case errors due to left out 
remainder terms are difficult 
to quantify. The continuum limit is in this way
only reached at an asymptotic rate proportional to $1/|\ln a|$
and can in practise not be taken.
Furthermore, rooted staggered quarks with their doubtful theoretical status are employed.
In spite of this, a phenomenological observation
adds support to the analysis of \cite{alpha:HPQCD}: there is 
agreement of results from several different variations of the theme.

The above mentioned difficulties have been a motivation for the 
ALPHA collaboration to work on a program which essentially is
based only on the assumptions that the continuum limit of the 
lattice theory exists and asymptotic freedom is present 
non-perturbatively. The resulting errors in $\alpha_\mrm{s}$ are
dominantly statistical and also the systematic component 
can be reduced further when the overall precision is improved in the future. 
In particular, as has been explained many times 
\cite{lat93:martin,reviews:leshouches2,nara:rainer},
\bi
  \vspace*{-1mm}\item $\alpha$ is defined non-perturbatively in a physical 
    (regularization independent) scheme,
  \vspace*{-1mm}\item this running can be efficiently computed numerically by a
    finite size strategy, 
  \vspace*{-1mm}\item the continuum limit can be taken in individual steps free of multi-scale problems,
  \vspace*{-1mm}\item renormalized perturbation theory is used only at large
    scales, where its precision is furthermore validated non-perturbatively.
\ei
\vspace*{-1mm}These properties come with a price.  Since a specific scheme
had to be devised in order to make precise non-perturbative computations 
possible, the $\beta$-function at 3-loop order needed to be
calculated~\cite{SF:LNWW,SF:stefan2,pert:2loopLW,pert:2loop_SU2,pert:2loop_nf0,pert:1loop,pert:2loopHP,pert:2loop_fin}. The same perturbative calculations are also
important to remove dominant parts of the discretization errors
(see section II.2 of \cite{nara:rainer}).
Furthermore a controlled non-perturbative running clearly
has to include the strange and charm quarks; this step has only been 
carried out recently and we report on it here. Still, as will be discussed in the conclusions
more work is needed before a numerical value for $\alpha_s(M_\mrm{Z})$ can
be given for phenomenology. It is also for these reasons that the
ALPHA collaboration has so far not been able to publish a value of 
$\alpha$ in the physical theory with all quarks which we would like to be
taken into account in the world average.

%%%%%%%%%%%%%%%%%%%%%%%%%%%%%%%%%%%%%%%%%%%%%%%%%%%%%%%%%%%%
\begin{wrapfigure}{r}{0.5\textwidth}
% \begin{figure}[htb!]
\vspace*{-0.8cm}
%\begin{center}
\includegraphics[width=0.52\textwidth]{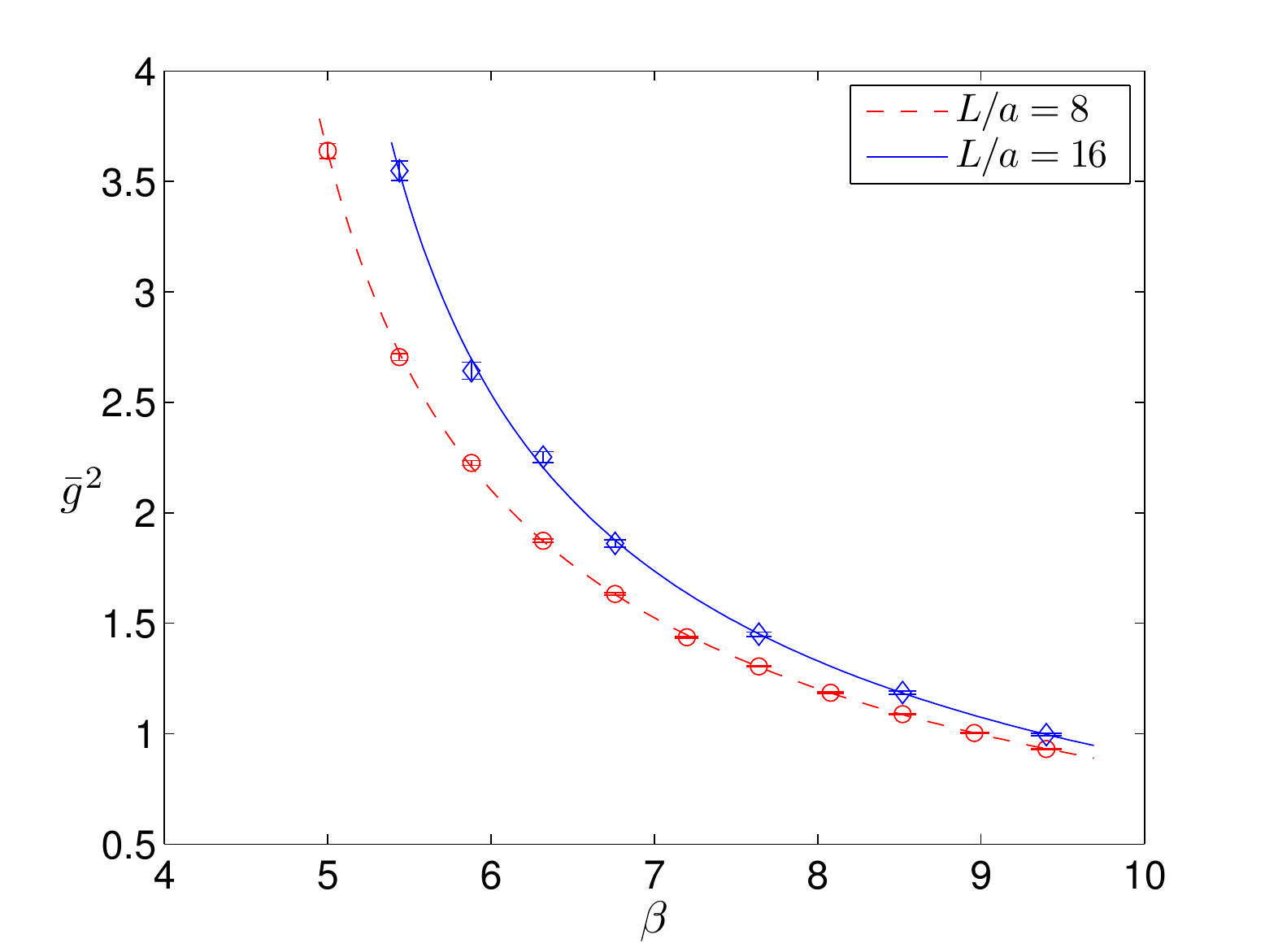}
%\end{center}
\vspace*{-0.7cm}
\caption[]{\label{f:gbarsqbeta} 
Interpolation \protect\eq{fitfunction}.
}
%\end{figure}
\end{wrapfigure}
%%%%%%%%%%%%%%%%%%%%%%%%%%%%%%%%%%%%%%%%%%%%%%%%%%%%%%%%%%%%

\section{Computation of the step scaling function}
In our finite size strategy, the coupling $\gbar^2$ is defined in a 
Euclidean space-time of size $L^4$ with \SF boundary conditions
\cite{SF:LNWW,SF:stefan2} and a 
renormalization scale $\mu=1/L$. The discrete scale evolution
defines the step scaling function $\sigma$ via
\begin{equation}\label{e:sigma}\nonumber
 \gbar^2(sL)=\sigma (s,\gbar^2(L))=\lim_{a/L\to 0} \Sigma (s,\gbar^2(L),a/L),
\end{equation}
which can be computed non-perturbatively as the continuum limit of
the lattice approximant $\Sigma$ as indicated. We chose a scale factor
$s=2$ for practical 
reasons and omit this argument from now on. 
In the Schr\"odinger functional, Dirichlet boundary conditions are imposed in Euclidean time
and therefore the $\rmO(a)$ Symanzik improvement of the theory requires boundary terms. Their 
coefficients are taken from perturbation theory, exactly as in 
\cite{Tekin:2009}, where we determined the crucial bulk $\rmO(a)$ improvement
coefficient $\csw$ non-perturbatively.

The computation of the lattice step scaling function, $\Sigma(u,a/L)$ requires to tune 
the bare mass and the bare coupling of the theory such that the PCAC mass 
(defined exactly as in \cite{alpha:nf2,alpha:nf4}) vanishes
and $\gbar^2(L)=u$. At the same bare parameters one then computes $\Sigma(u,a/L)=\gbar^2(2L)$.
An explicit 2-dimensional tuning is rather cumbersome. We therefore followed
\cite{alpha:yale}, picked a series of bare couplings $g_0^2=6/\beta$ 
and tuned the PCAC mass to zero. For those bare parameters 
we then compute $\bar{g}^2(\beta,L/a)$ 
and $\bar{g}^2(\beta,2L/a)$ and 
interpolated to the desired values of $u$ via
%%%%%%%%%%%%%%%%%%%%%%%%%%%%%%%%%%%%%%%%%%%%%%%%%%%%%%%%%%%%
\begin{wrapfigure}{r}{0.5\textwidth}
\vspace*{-0.0cm}
\includegraphics[width=0.5\textwidth]{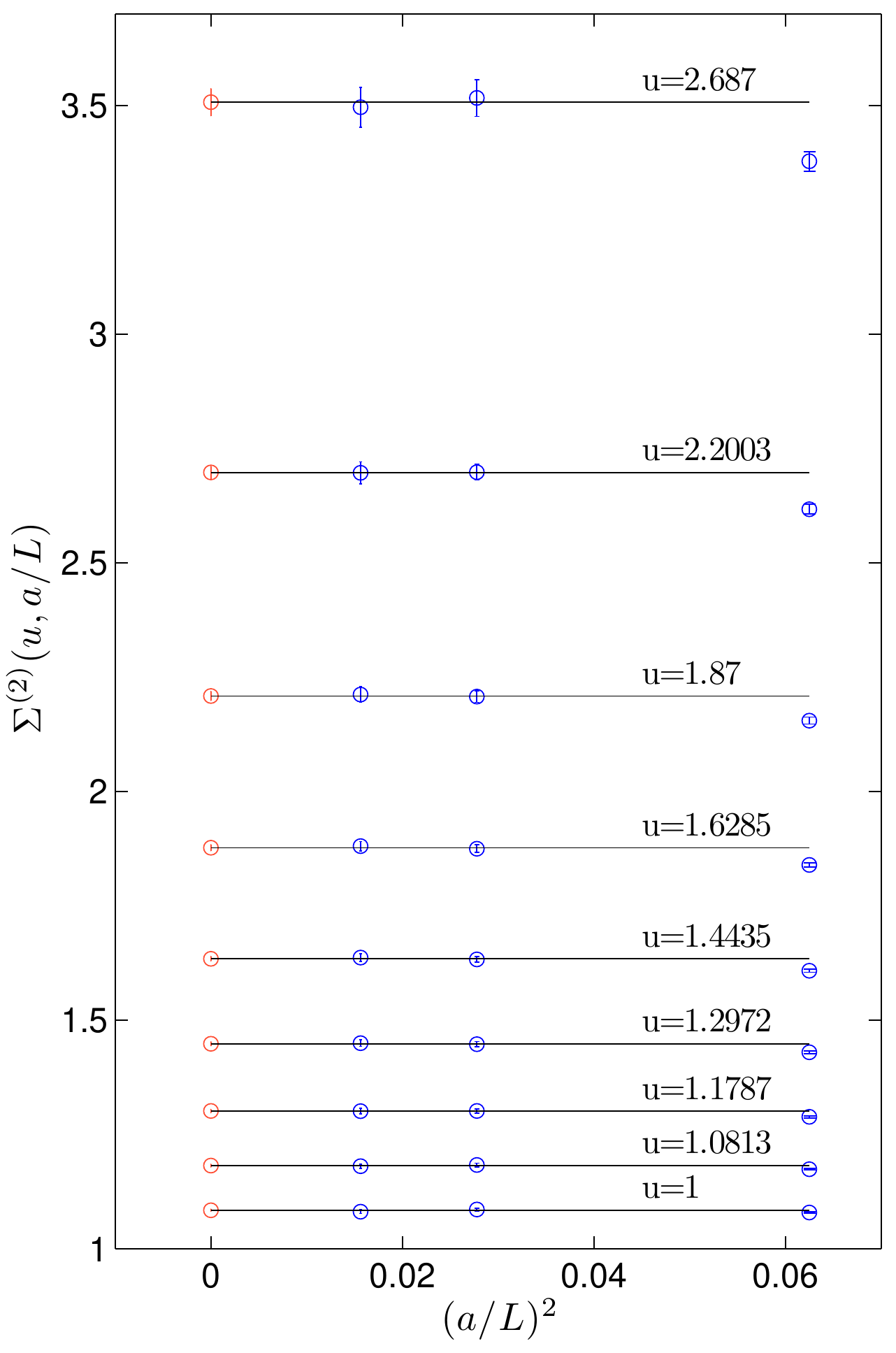}
\vspace*{-3mm}
\caption[]{\label{f:continuum} 
Constant fit continuum extrapolation.
}
\vspace*{6mm}
\end{wrapfigure}
%%%%%%%%%%%%%%%%%%%%%%%%%%%%%%%%%%%%%%%%%%%%%%%%%%%%%%%%%%%%
\begin{equation}\label{fitfunction}
\bar{g}^2(\beta,L/a)=\frac{6}{\beta}\left[\sum_{m=0}^{M}c_{m,L/a}\left(\frac{6}{\beta}\right)^m\right]^{-1}\,
\end{equation}
motivated by perturbation theory. We do not fix the 
known perturbative expansion coefficients, not even 
$c_{0,L/a}$. The interpolations for $L/a=8$ are illustrated
in \fig{f:gbarsqbeta}. $\Sigma(u,1/8)$ is given by the value of the
upper curve at the $\beta$ where the lower one passes through
$\bar{g}^2(\beta,8)=u$. Stability of the interpolations
with respect to $M$ was checked.

Using $\Sigma$ from the interpolation, we form the 
2-loop improved 
lattice step scaling function~\cite{alpha:SU2impr} 
\begin{equation}
 \Sigma^{(2)}(u,a/L)=\frac{\Sigma (u,a/L)}{1+\delta_{1}(a/L)u+\delta_{2}(a/L)u^2}
\end{equation}
with $\delta_1,\delta_2$ known from 
\cite{alpha:SU3,pert:2loop_nf0,pert:1loop,pert:2loopHP,pert:2loop_fin}.
We expect $\Sigma^{(2)}$ to 
have smaller overall cutoff effects. Asymptotically, they still start at order
$a\times u^4$ but terms of order $a^m\times u^n$ are removed for
all $m$ and for $n\leq3$ (in fact non-perturbatively in $a$). 
As mentioned previously, the order $a\times u^4$ terms are due to the 
only perturbatively known boundary improvement terms. Their influence
was explicitly checked for $\nf=2$ and found to be minor \cite{alpha:nf2,nara:rainer}
for our action, 
at least when $\ct$ is known to 2-loop order.
We therefore assume that  the step scaling function 
converges {\em effectively} at a rate
\begin{equation}
  \Sigma^{(2)}(u,a/L)=\sigma(u)+\rmO(a^2).
\end{equation}
To study the continuum limit and its uncertainty 
we carried out three different analysis.
\begin{itemize}
 \vspace*{-1mm}\item {\em Constant fit:} A fit of $\Sigma^{(2)}(u,a/L)$ for $L/a=6,8$ to a constant, for
      each $u$. 
 \vspace*{-1mm}\item {\em Global fit:} A fit
  %\begin{equation} \label{imprsigma}
  $  \Sigma^{(2)}(u,a/L)=\sigma(u)+\rho\, u^4\,(a/L)^2\,, $
  %\end{equation}
  with a separate, independent parameter $\sigma(u)$ for
  each value $u$ but a common parameter $\rho$ modelling the cutoff-effects. 
 \vspace*{-1mm}\item {\em $L/a=8$ data:} Using directly $\sigma(u)=\Sigma^{(2)}(u,1/8)$.
\end{itemize} 
\vspace*{-1mm}The three different ans\"atze yield results which are in complete agreement with each other. 
The global fit returns $\rho=0.007(85)$ which is a good indication that cutoff effects 
are negligible in the data for $L/a=6,8$. In order to have a safe error estimate 
on the continuum limit
we chose just the $L/a=8$ data as our present result.

\section{The running of the coupling}
\begin{wrapfigure}{r}{0.63\textwidth}
%\begin{figure}[htb!]
\vspace*{-2.0cm}
\begin{center}
\includegraphics[width=0.63\textwidth]{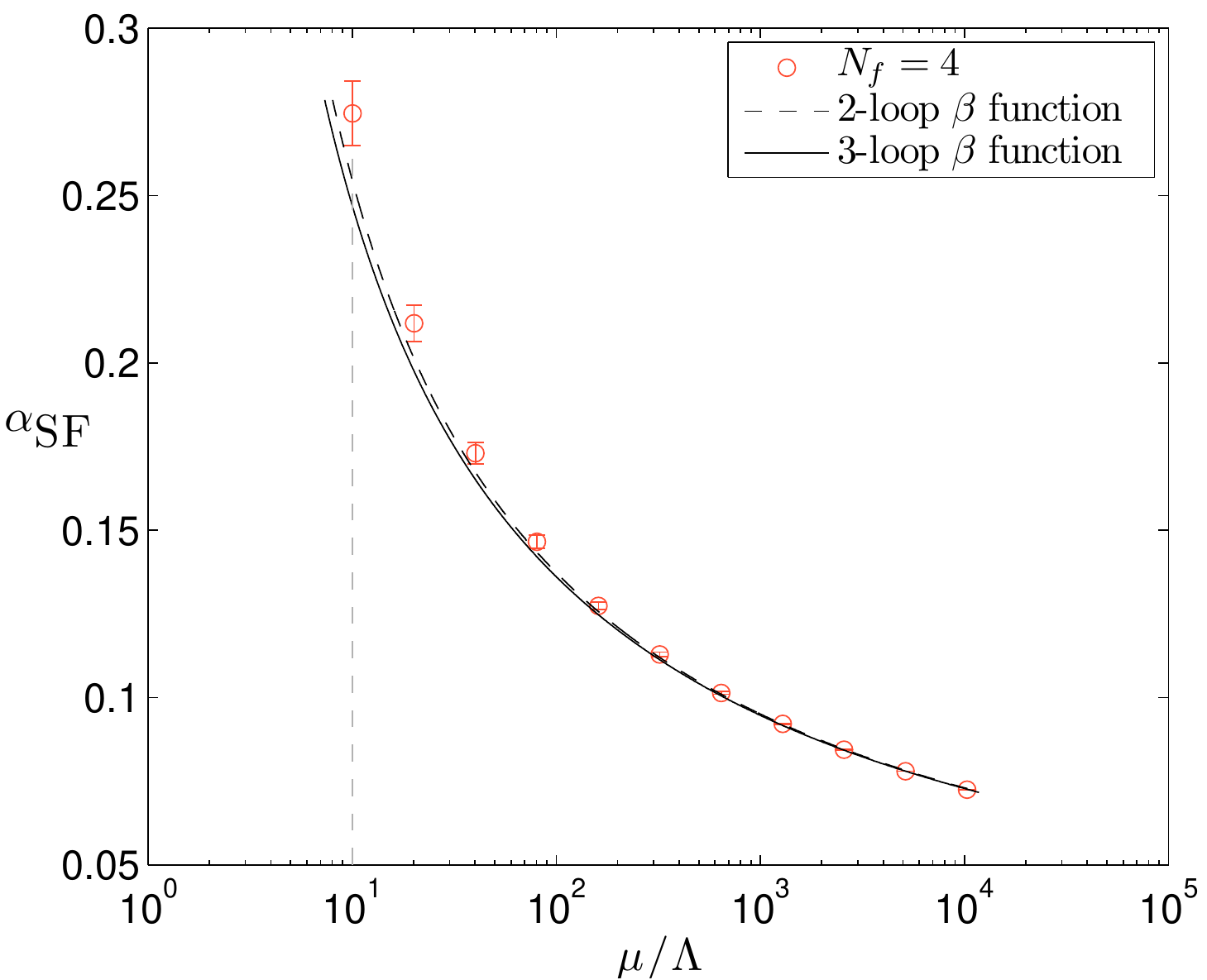}
\end{center}
\vspace*{-0.5cm}
\caption[]{\label{f:running} 
Non-perturbative running coupling compared to perturbation theory.
}
%\end{figure}
\end{wrapfigure}

A polynomial interpolation 
$\sigma(u)=u+s_{0}u^2+s_{1}u^3+0.0036\,u^4-0.0005\,u^5,$ with 
the coefficients up to $u^3$ fixed by perturbation theory
represents $\sigma(u)$ in the range $0.9\leq u \leq2.7$ with negligible
interpolation errors.
The running of the coupling 
is then obtained by solving the recurrence 
\begin{equation}\label{recursion}
 u_{i}=\sigma(u_{i+1}),\quad i=0,\dots,n,\quad u_{0}=\umax=\gbar^2\left(\Lmax\right)\,
\end{equation}
for $u_i=\gbar^2(L_i)$, $L_i= 2^{-i}\Lmax$.
As \fig{f:running} shows, agreement with perturbative running
at the 3-loop level is found at the highest scales in agreement with standard 
estimates of remainder terms. This allows us to relate $u_{9}$ to $L_9 \times \Lambda$
by using the 3-loop $\beta$-function [for couplings up to $u_9$ only].
Then with the
non-perturbative $\sigma(u)$ used in \eq{recursion} 
we can connect to
larger values of $u$,
for example
\begin{equation}\label{lnlambdalmax}
 \ln(\Lambda\Lmax)=-2.294(83)\quad\text{ at }\gbar^2(\Lmax)=\umax=3.45\,.
\end{equation}
While a precise MeV value still has to be determined we clearly expect $\Lmax$ to lie
in the range of hadronic scales.
The (uncorrelated) errors of our primary MC data for $\gbar^2(\beta,L/a)$
are propagated through all steps of the analysis as described in
\cite{alpha:nf4}. Note that some steps, in particular \eq{recursion},
introduce correlations into the final results. The errors of the
points \fig{f:running} are therefore not independent.

%%%%%%%%%%%%%%%%%%%%%%%%%%%%%%%%%%%%%%%%%%%%%%%%%%%%%%%%%%%%
%%%%%%%%%%%%%%%%%%%%%%%%%%%%%%%%%%%%%%%%%%%%%%%%%%%%%%%%%%%%

\section{Conclusions}

At the present level of statistical errors, systematic errors from the continuum
extrapolation are almost certainly negligible.
In the future, in particular with
further reduced statistical errors, data with
larger $L/a$ clearly have to be added.
It will also be interesting to compare the efficiency of computations
with different regularizations of the Schr\"odinger functional, such as 
chirally rotated boundary conditions \cite{SF:chirrot2,lat09:jeni}
and staggered quarks \cite{lat07:paula,lat10:paula}.

We do observe a small but significant deviation from 
3-loop perturbation theory at the largest coupling reached in \fig{f:running}. 
It is about 10\% (three standard deviations) and the \SF coupling has a value of 
$\alphaSF \approx 0.28$. For $\nf=2$ a similar effect was visible only for 
larger coupling~\cite{alpha:nf2}. These findings underline the necessity of 
going to weak coupling before applying perturbation theory. 

We are now a good step closer to the computation
of the $\Lambda$-parameter in 4-flavor QCD, which may then be perturbatively connected to 
e.g. the 5-flavor $\alphamsbar(M_\mrm{Z})$. 
However the low energy scale $\Lmax$ that was introduced for technical convenience, remains to be expressed in 
physical units
through large volume 4-flavor simulations and we may want to improve the 
precision in \fig{f:running}.

\noindent
{\bf Acknowledgements.}
We thank NIC and DESY for allocating computer time on the APE
computers to this project and the APE group for its help. This
work is supported by the  Deutsche Forschungsgemeinschaft
in the SFB/TR~09 and by
the European community through
EU Contract No.~MRTN-CT-2006-035482, ``FLAVIAnet''.

\bibliographystyle{JHEP}   %if you use h-elsevier.bst
\bibliography{latticen,qcd}           %or whatever your .bib file i

%\begin{thebibliography}{99}
%\bibitem{...} 
%....
%\end{thebibliography}

\end{document}